\documentclass[conference]{IEEEtran}
\IEEEoverridecommandlockouts
\usepackage{cite}
\usepackage{booktabs}
\usepackage{amsmath,amssymb,amsfonts}
\usepackage{algorithm}%
\usepackage{algorithmicx}%
\usepackage{algpseudocode}%
\usepackage{makecell} 
\usepackage{graphicx}
\usepackage{textcomp}
\usepackage{xcolor}
\usepackage{cuted}
\usepackage{stfloats}
\usepackage{bm} 
\def\BibTeX{{\rm B\kern-.05em{\sc i\kern-.025em b}\kern-.08em
    T\kern-.2em\lower.7ex\hbox{E}\kern-.125emX}}
\begin{document}

\title{Joint Beamforming Design and Port Selection in Fluid Antenna-Assisted Multi-Cell Networks: A Personalized Federated Learning Approach
\thanks{This work is supported in part by the Scientific Research Program of Shaanxi Provincial Education Department under Grant 24JK0674 and Natural Science Foundation of Shaanxi Province under Grant 2025JC-YBQN-889. (\emph{Corresponding author: Xing Hao})}
}

\author{
    \IEEEauthorblockN{
        Liwen Gao\textsuperscript{1}, 
          Li Zheng\textsuperscript{2},
        Xing Hao\textsuperscript{1*},
        Ziru Chen\textsuperscript{3}, 
        Lin X. Cai\textsuperscript{3}
    }
    
    \vspace{0.15cm} 
    
    \IEEEauthorblockA{\textsuperscript{1}School of Electronic Information, \textsuperscript{2}College of Computer Science, Northwest University, Xi'an, China}
    
    
    \IEEEauthorblockA{\textsuperscript{3}Department of Electrical and Computer Engineering, Illinois Institute of Technology, Chicago, USA}

    
    \IEEEauthorblockA{\{gaoliwen,lzheng\}@stumail.nwu.edu.cn, *xhao@nwu.edu.cn, zchen@ofinno.com, 
lincai@ieee.org}   
}

\maketitle
\thispagestyle{empty}
\pagestyle{empty}
\begin{abstract}
This paper investigates joint beamforming and port selection in multi-cell fluid antenna-assisted (FAS) networks. In such networks, active beamforming and discrete FA port selection are coupled through intra-cell and inter-cell interference and are jointly optimized to maximize the weighted sum-rate (WSR). We develop a federated representation learning (FedRep) framework with a position-aware dual-branch deep neural network (PA-DNN). The PA-DNN uses channel state information and port positional encoding as inputs, and jointly outputs beamforming vectors and port selections through two task-specific branches. To support decentralized training across heterogeneous cells, the FedRep framework shares global beamforming-related parameters among base stations while keeping port-selection parameters local for cell-specific adaptation. Simulation results show that the proposed scheme achieves a higher weighted sum-rate than conventional FL and port-selection benchmark schemes.

\end{abstract}


\section{Introduction}
The rapid evolution of sixth-generation (6G) wireless networks is expected to support ultra-high data rates, massive connectivity, and highly adaptive radio environments. To meet these requirements, reconfigurable wireless technologies have attracted increasing attention by providing additional spatial degrees of freedom beyond conventional fixed-position antenna (FPA)~\cite{fa2}. Among them, the fluid antenna (FA) has recently emerged as a promising candidate technique. Different from conventional FPAs, FASs can dynamically switch or adjust their effective radiating position within a compact spatial region, thereby exploiting spatial channel diversity to improve link reliability and mitigate interference~\cite{x,fa5}. 

Several studies have investigated port selection in fluid antenna-assisted wireless communication~\cite{port2,portselect}.
Some works analyzed the outage probability and diversity gain, demonstrating the potential of FAS in improving link reliability \cite{port2}. Others proposed algorithms that select the best port from a few observations by leveraging the Bessel-function-based spatial correlation, demonstrating the feasibility and efficiency of utilizing spatial correlation for port selection in FAS \cite{portselect}. These studies show that selecting a proper port is essential for exploiting the spatial diversity of FAS and improving the received signal quality. 

Some studies have extended FAS optimization to multi-user transmission~\cite{11,joint2}. To address the existing intra-cell interference, \cite{11} proposed a convolutional neural network (CNN)-based method that captures local spatial correlations among ports via convolutional kernels sliding over the spatial sequence, enabling it to extrapolate channels at unobserved ports from a few observed ones and then select the best port to avoid interference. Meanwhile, most works jointly optimized coupled active beamforming and port selection. Specifically, \cite{joint2} develops a two-stage graph neural network (GNN)-based approach to jointly infer fluid antenna positions and beamforming matrices by capturing the interference topology among users.
Most existing multi-user FAS studies are limited to single-cell scenarios. In practice, dense base station (BS) deployment introduces severe inter-cell interference, which tightly couples the beamforming and port selection variables across adjacent cells, making single-cell methods inapplicable. Therefore, investigating joint optimization in multi-cell FAS environments is critical.

The work~\cite{RDL} proposed a reinforcement learning-based approach where each BS acts as an independent agent, using a world model for dynamics prediction, a Pointerformer for port selection,  and zero-forcing beamforming with power prediction. The work~\cite{multi} studied a downlink FAS network where FA port selection and active beamforming are jointly optimized to maximize the weighted sum-rate(WSR).
However, both \cite{RDL} and \cite{multi} rely on centralized training, which may fails to accommodate per-BS preferences such as user-specific weights in a weighted sum-rate objective.

Federated learning (FL) has been widely used as a decentralized learning paradigm to reduce raw channel state information (CSI) aggregation at a central server and support local model adaptation~\cite{chen2024optimizing,ref}. However, conventional FL typically learns a single global model for all BSs, which may ignore local data diversity under heterogeneous multi-cell conditions. To improve personalization, pFedMe{~\cite{pfedme}} introduces an $L_2$ regularizer into local losses, while EM-based personalized federated learning (PFL)~\cite{ni2026pfedwn} adopts adaptive aggregation weights. Federated representation learning (Fedrep)~\cite{fedrep} further decouples model parameters into a globally shared representation and locally personalized heads. However, existing PFL and FedRep-based methods have not been tailored to the coupled active beamforming and FA port selection problem in multi-cell FAS networks. To the best of our knowledge, FL-based joint optimization of active beamforming and FA port selection in multi-cell FAS networks remains unexplored.

In this paper, we maximize the WSR in a multi-cell network by jointly optimizing BS active beamforming and users' fluid-antenna port selection under power constraints. First, to enable decentralized training under per-cell heterogeneity, we adopt the FedRep framework, where FA port selection is treated as the personalized local head since it is a receive-side choice that can be made locally, while beamforming is assigned to the globally shared representation since its active transmission direction inherently affects neighboring cells and requires cross-cell coordination. Second, we propose a position-aware dual-branch DNN (PA-DNN) for the resulting mixed-integer problem, which fuses raw CSI with physical positional encoding to capture spatial correlations across arbitrary port arrangements. The shared features are fed into beamforming and port-selection branches, with Gumbel-Softmax enabling differentiable discrete port selection. Simulation results confirm the superiority of the proposed scheme over benchmark approaches.

\section{System Model and Problem Formulation}
We consider a downlink multi-cell communication network with $M$ cells, where each cell consists of a BS equipped with $N$ fixed-position antennas serving $K$ users. Each user is equipped with a fluid antenna comprising $P$ ports that can be switched within a two-dimensional square region. Let $\mathcal{P} = \{ 1 ,2 ,\dots,  P  \}$ denote the set of available FA port indices, and  $\mathcal{C} = \{c_{m,k}   \}_{\forall m,k}$ denote the set of activated port indices, where $c_{m,k} \in \mathcal{P}$ is the selected port of user $k$ in cell $m$. To describe both desired and interfering links, we use $i \in \{1,2,\dots,M\}$ to denote the index of an arbitrary transmitting BS, while $m \in \{1,2,\dots,M\}$ denotes the index of the cell where the considered user is located.

\subsection{Channel Model}
We establish local three-dimensional (3D) Cartesian coordinate systems for each BS and user to describe the positions of the fixed antennas and FA ports relative to their respective local origins.

The position of the $n$-th transmit antenna at BS $i$ is denoted by  $\mathbf{b}_{i,n}=[x^B_{i,n},y^B_{i,n},z^B_{i,n}]^{T},  1\leq n\leq N$.
 The position of the  $p$-th port of user $k$ in cell $m$ is denoted by $\mathbf{q}_{m,k,p}=[x^U_{m,k,p},y^U_{m,k,p},z^U_{m,k,p}]^{T},  1\leq p\leq P.$

Each BS-user link is modeled with $L$ propagation paths. For the link from BS $i$ to user $k$ in cell $m$, the complex gain of the $\ell$-th path is denoted by $\alpha_{i,m,k,\ell}\sim\mathcal{CN}(0,1)$, $\ell\in\{1,\ldots,L\}$. Accordingly, the path-response matrix is given by
\begin{equation}
\mathbf{\Sigma}_{i,m,k} = \operatorname{diag}  \left( \alpha_{i,m,k,1}, \ldots, \alpha_{i,m,k,L}  \right).
\label{eq:PRM}
\end{equation}

Without loss of generality, we consider the channel between BS $i$ to user $k$ in cell $m$. To simplify notation, the link indices $(i,m,k)$ are omitted in the definitions of the phase differences and field-response vectors (FRVs). Specifically,
the receive  and transmit phase differences of the $\ell$-th path are respectively given by
\begin{equation}
\begin{aligned}
&\quad \rho^r_{\ell}(\mathbf{q}_{m,k,p})\\
&= x^{\rm U}_{m,k,p}\cos\phi^r_{\ell}\cos\theta^r_{\ell}
+y^{\rm U}_{m,k,p}\cos\phi^r_{\ell}\sin\theta^r_{\ell}
+z^{\rm U}_{m,k,p}\sin\phi^r_{\ell},
\end{aligned}
\label{eq:receive_phase}
\end{equation}

\begin{equation}
\begin{aligned}
\rho^t_{\ell}(\mathbf{b}_{i,n})
=& x^{\rm B}_{i,n}\cos\phi^t_{\ell}\cos\theta^t_{\ell}
+y^{\rm B}_{i,n}\cos\phi^t_{\ell}\sin\theta^t_{\ell}
+z^{\rm B}_{i,n}\sin\phi^t_{\ell},
\end{aligned}
\label{eq:transmit_phase}
\end{equation}
where $\theta^r_{\ell}$ and $\phi^r_{\ell}$ denote the azimuth and elevation angles of arrival, respectively, and $\theta^t_{\ell}$ and $\phi^t_{\ell}$ denote the azimuth and elevation angles of departure, respectively.

The receive and transmit FRVs for the channel between the BS $i$ and the user $k$ in cell $m$ are respectively written as

\begin{equation}
\begin{aligned}
& \quad \mathbf{g}_k(\mathbf{q}_{m,k,p})\\
&= \big[
 e^{j\frac{2\pi}{\lambda}\rho^r_{1}(\mathbf{q}_{m,k,p})},
e^{j\frac{2\pi}{\lambda}\rho^r_{2}(\mathbf{q}_{m,k,p})},
\ldots,e^{j\frac{2\pi}{\lambda}\rho^r_{L}(\mathbf{q}_{m,k,p})}
\big]^T,
\end{aligned}
\label{eq:receive_frv}
\end{equation}

\begin{equation}
\begin{aligned}
\mathbf{f}_k(\mathbf{b}_{i,n})
=
\big[
& e^{j\frac{2\pi}{\lambda}\rho^t_{1}(\mathbf{b}_{i,n})},
e^{j\frac{2\pi}{\lambda}\rho^t_{2}(\mathbf{b}_{i,n})},
\ldots,
e^{j\frac{2\pi}{\lambda}\rho^t_{L}(\mathbf{b}_{i,n})}
\big]^T,
\end{aligned}
\label{eq:transmit_frv}
\end{equation}
where $\lambda$ denotes the carrier wavelength. Since the transmit antennas at each BS are fixed-position antennas,
the transmit field-response matrix for the channel between BS $i$ and user $k$ in cell $m$ is given by
\begin{equation}
\mathbf{F}_{i,m,k}
=
\left[
\mathbf{f}_k(\mathbf{b}_{i,1}),\mathbf{f}_k(\mathbf{b}_{i,2}),
\ldots,
\mathbf{f}_k(\mathbf{b}_{i,N})
\right]
.
\label{eq:FRM}
\end{equation}

The channel vector between BS $i$ and user $k$ in cell $m$ at the $p$-th FA port is expressed as
\begin{equation}
\mathbf{h}_{i,m,k}(\mathbf{q}_{m,k,p})
=
\mathbf{F}_{i,m,k}^{H}  \mathbf{\Sigma}_{i,m,k}  \mathbf{g}(\mathbf{q}_{m,k,p}).
\label{eq:channel_vector}
\end{equation}
In this equation, $\mathbf{h}_{m,m,k}(\mathbf{q}_{m,k,p})$ represents the channel from the serving BS to user $k$ in cell $m$, while $\mathbf{h}_{i,m,k}(\mathbf{q}_{m,k,p})$ represents the inter-cell interfering channel from BS $i$ to this user.
\subsection{System Model}
Let $\mathbf{X}_m \in \mathbb{C}^{N \times L}$ be the transmitted signal matrix from the BS in cell $m$, with $L$ the frame length. The transmitted signal is

\begin{equation}
\mathbf{X}_m = \mathbf{W}_m\mathbf{S}_m = \sum_{k=1}^{K} \mathbf{w}_{m,k}\mathbf{s}_{m,k},
\label{eq:transmit_signal}
\end{equation}
where $\mathbf{W}_m = [\mathbf{w}_{m,1},\dots,\mathbf{w}_{m,K}] \in \mathbb{C}^{N \times K}$ is the beamforming matrix, $\mathbf{S}_m \in \mathbb{C}^{K \times L}$ is the per-cell symbol matrix with $k$-th row $\mathbf{s}_{m,k} \in \mathbb{C}^{1\times L}$ being the data stream for user $k$.

We assume that each user can only activate a single port for data reception at any given time. Let $\mathbf{p}_{m,k} \in \mathcal{P}$ denote the specific position of the activated port for user $k$ in cell $m$. The received signal vector \(\mathbf{y}_{m,k}^H \) at user \(k\) in cell \(m\) is
\begin{equation}
\begin{split}
\mathbf{y}_{m,k}^H =& \;\mathbf{h}_{m,m,k}^H \mathbf{w}_{m,k}\mathbf{s}_{m,k}
   + \sum_{j\neq k} \mathbf{h}_{m,m,k}^H \mathbf{w}_{m,j}\mathbf{s}_{m,j} \\
   & + \sum_{i\neq m}\sum_{j=1}^K \mathbf{h}_{i,m,k}^H \mathbf{w}_{i,j}\mathbf{s}_{i,j}
   + \mathbf{n}_{m,k}^H,
\end{split}
\label{eq:received_signal}
\end{equation}
where 
\(\mathbf{n}_{m,k} \sim \mathcal{CN}(\mathbf{0}, \sigma_c^2 \mathbf{I}) \) is the additive white Gaussian noise  vector. The signal-to-interference-plus-noise ratio (SINR) at user $k$ in cell $m$ is given as \eqref{eq:sinr_comm}  at the bottom of this page.
\begin{figure*}[b]
\hrulefill
\begin{equation}
\begin{aligned}
 \zeta_{m,k} = \frac{\left| \mathbf{h}_{m,m,k}^H(\mathbf{q}_{m,k,p})\mathbf{w}_{m,k} \right|^2}{\sum_{j \neq k}^{K} \left| \mathbf{h}_{m,m,k}^H(\mathbf{q}_{m,k,p})\mathbf{w}_{m,j} \right|^2 + \sum_{i \neq m}^{M} \sum_{j=1}^{K} \left| \mathbf{h}_{i,m,k}^H(\mathbf{q}_{m,k,p})\mathbf{w}_{i,j} \right|^2 + \sigma_c^2}. 
\label{eq:sinr_comm}
\end{aligned}
\end{equation}
\end{figure*}
Thus, the achievable sum rate for user $k$ in cell $m$ is given by
\begin{equation}
R_{m,k}=\log_2\left(1+\zeta_{m,k}\right).
\label{eq:sumrate_comm}
\end{equation}

\subsection{Problem Formulation}

Let $\mathcal{W} = \{\mathbf{W}_m, \forall m\}$ denote the set of beamforming matrices. We formulate the weighted sum-rate maximization problem as
\begin{subequations}
\begin{align}
\max_{\mathcal{W}, \mathcal{C} } \quad & \sum_{m=1}^{M} \sum_{k=1}^{K} \omega_{m,k} R_{m,k} \\
\text{s.t.} \quad & \|\mathbf{W}_m\|_F^2 \leq P_T, \quad \forall m, \label{power}\\
& c_{m,k} \in \mathcal{P}, \quad \forall m,k,\label{port}
\end{align}
\label{eq:opt_problem}
\end{subequations}
where $\omega_{m,k}\geq 0$ denotes the priority weight of user $k$ in cell $m$, and $P_T$ is the maximum transmit power of each BS. The first constraint \eqref{power} denotes the transmit power budget of each BS, and the second constraint \eqref{port} indicates that each user selects one activated FA port from the available port set~$\mathcal{P}$.  

\section{Proposed Solution}
In this section, we first present the PA‑DNN architecture, followed by the FedRep‑based training framework.

\subsection{PA-DNN-based beamforming design and port selection}
\subsubsection{Input and Output}
We design the input as two parts, the CSI matrix and a positional encoding (PE) that endows spatial awareness. For a specific cell $m$, the CSI matrix $\mathbf{X}_m^{\mathrm{CSI}} \in \mathbb{R}^{KP \times 2}$ is formed by separating the real and imaginary parts of channel gains for all $K$ users across all $P$ ports, which is explicitly formulated as
\begin{equation}
\mathbf{X}_m^{\mathrm{CSI}} = \begin{bmatrix}
\mathrm{Re}\{h_{m,1,1}\} & \mathrm{Im}\{h_{m,1,1}\} \\
\vdots & \vdots \\
\mathrm{Re}\{h_{m,K,P}\} & \mathrm{Im}\{h_{m,K,P}\}
\end{bmatrix}.
\label{eq:CSI_input}
\end{equation}
Specifically, this PE projects the physical 3D coordinates into a high-dimensional space via multi-frequency sine and cosine functions, the positional encoding (PE) matrix can be represented as:
\begin{equation}
\begin{split}
\mathbf{e}_{m,k,p} = \Big[ & \sin(x_{m,k,p}^{U}\omega_1), \cos(x_{m,k,p}^{U}\omega_1), \dots, \\
& \sin(x_{m,k,p}^{U}\omega_D), \cos(x_{m,k,p}^{U}\omega_D), \\
& \sin(y_{m,k,p}^{U}\omega_1), \cos(y_{m,k,p}^{U}\omega_1), \dots, \\
& \sin(y_{m,k,p}^{U}\omega_D), \cos(y_{m,k,p}^{U}\omega_D), \\
& \sin(z_{m,k,p}^{U}\omega_1), \cos(z_{m,k,p}^{U}\omega_1), \dots, \\
& \sin(z_{m,k,p}^{U}\omega_D), \cos(z_{m,k,p}^{U}\omega_D) \Big]^T
\end{split}
\label{eq:PE_vector}
\end{equation}
where $\boldsymbol{\omega} = [\omega_1, \dots, \omega_{D}]$ denotes a set of multi-scale frequency components. By mathematically mapping relative spatial distances into measurable feature distances through these varying frequencies, this continuous transformation enables the network to naturally learn that closely spaced ports experience similar channel fading. Therefore, the comprehensive PE matrix for all $K$ users in cell $m$ can be expressed as
\begin{equation}
\mathbf{X}_m^{\mathrm{PE}} = \begin{bmatrix}
\mathbf{e}_{m,1,1}^T \\
\vdots \\
\mathbf{e}_{m,1,P}^T \\
\vdots \\
\mathbf{e}_{m,K,1}^T \\
\vdots \\
\mathbf{e}_{m,K,P}^T
\end{bmatrix} \in \mathbb{R}^{KP \times 6D}.
\label{eq:PE_matrix}
\end{equation}
Specifically, the network input is constructed by fusing the CSI matrix and the PE matrix, given by
\begin{equation}
\mathbf{X}_m^{\mathrm{in}} = \left[\mathbf{X}_m^{\mathrm{CSI}}, \mathbf{X}_m^{\mathrm{PE}}\right] \in \mathbb{R}^{KP \times (2 + 6D)}.
\label{eq:joint_input}
\end{equation}
The network outputs the beamforming matrix and the port selections in parallel. The unnormalized beamforming matrix generated by the beamforming branch can be represented as
\begin{equation}
\hat{\mathbf{W}}_m^{\mathrm{out}} = \begin{bmatrix} 
\operatorname{Re}\{\hat{\mathbf{w}}_{m,1}\} & \operatorname{Im}\{\hat{\mathbf{w}}_{m,1}\} \\ 
\vdots & \vdots \\ 
\operatorname{Re}\{\hat{\mathbf{w}}_{m,K}\} & \operatorname{Im}\{\hat{\mathbf{w}}_{m,K}\} 
\end{bmatrix} \in \mathbb{R}^{N_T K \times 2},
\label{eq:beamforming_output}
\end{equation}
meanwhile, the following index set generated by the port selection branch can be represented as 
\begin{equation}
\mathcal{C}_m = \{ \mathcal{C}_{m,1}, \mathcal{C}_{m,2}, \dots, \mathcal{C}_{m,K} \}.
\label{eq:port_output}
\end{equation}

\subsubsection{Neural Network Architecture}
The joint input \(\mathbf{X}_m^{\mathrm{in}}\) is first processed by a shared backbone, which is a feature extractor composed of a multi-layer perceptron (MLP) with  hidden layers. To effectively extract the high-dimensional spatial information embedded in $\mathbf{X}_m^{\mathrm{in}}$, we utilize Sinusoidal Representation Networks (SIREN) in the hidden layers as a replacement for traditional ReLU activations. This design choice is primarily motivated by the periodic nature of the sine function, which prevents the high-frequency spatial features introduced by positional encoding from decaying or over-smoothing in deeper layers, thereby preserving the ability to distinguish different ports. 

Specifically, the backbone outputs a unified feature vector \(\mathbf{z}_m^{\mathrm{shared}}(\mathbf{X}_m^{\mathrm{in}})\). This latent representation is then processed by two task-specific branches in parallel.

The continuous beamforming branch takes \(\mathbf{z}_m^{\mathrm{shared}}\) and produces the unnormalized matrix \(\hat{\mathbf{W}}_m^{\mathrm{out}} = \mathrm{MLP1}(\mathbf{z}_m^{\mathrm{shared}})\) through several fully connected layers. A deterministic power projection layer is subsequently applied to enforce the per-BS power constraint, which can be expressed as
\begin{equation}
\mathbf{w}_{m,k} = \sqrt{P_T}\frac{\hat{\mathbf{w}}_{m,k}}{\sqrt{\sum_{j=1}^{K}\|\hat{\mathbf{w}}_{m,j}\|^2}}, \quad \forall k.
\label{eq:power_projection}
\end{equation}

Concurrently, the discrete port-selection branch maps \(\mathbf{z}_m^{\mathrm{shared}}\) via \(\mathrm{MLP2}\) to output a logit vector \(\boldsymbol{\pi}_{m,k} = [\pi_{m,k,1},\dots,\pi_{m,k,P}]^T\) for each user \(k\). The chosen port index \(\mathcal{C}_{m,k}\) is determined via the \(\arg\max\) operation, which converts the logit vector into a one-hot vector (e.g., \([0,\dots,1,\dots,0]^T\)). Consequently, the equivalent channel for user \(k\) is the response of the activated port, denoted as \(\mathbf{h}_{m,k}(\mathcal{C}_{m,k})\). Since the \(\arg\max\) operation is non-differentiable, we employ the Gumbel-Softmax reparameterization with a Straight-Through Estimator (STE) to yield \(\mathcal{C}_{m,k} = \text{STE-Gumbel}(\mathrm{MLP2}(\mathbf{z}_m^{\mathrm{shared}}))\). In this mechanism, the forward pass uses hard one-hot selection for exact port switching, while the backward pass uses the continuous Gumbel-Softmax relaxation for gradient computation. The \(p\)-th element of the relaxed output is given by
\begin{equation}
y_{m,k,p} = \frac{\exp((\pi_{m,k,p} + g_p)/\tau)}{\sum_{i=1}^{P}\exp((\pi_{m,k,i} + g_i)/\tau)},
\label{eq:gumbel_softmax}
\end{equation}
where \(g_p \sim \text{Gumbel}(0,1)\) and \(\tau > 0\) is the temperature parameter. This dual-branch design ensures end-to-end differentiability while resolving the inherent mixed-integer constraints.

\subsection{FedRep-Based PFL}
Under the FedRep paradigm, the model structure is divided into two functional modules: the global representation and the local head. Accordingly, the overall trainable parameters of the PA-DNN at BS \(m\), denoted as \(\Theta_m\), are correspondingly decoupled into a globally shared set \(\Theta_m^{\mathrm{global}}\) and a locally personalized set \(\Theta_m^{\mathrm{local}}\) as follows:
\begin{equation}
    \Theta_m = \left\{ \Theta_m^{\mathrm{global}}, \Theta_m^{\mathrm{local}} \right\}.
    \label{eq:parameter_decoupling}
\end{equation}
The global parameters $\Theta_m^{\mathrm{global}}$, which correspond to the beamforming branch, are shared among base stations to learn universal spatial features common to all cells, thereby improving overall generalization. In contrast, the local parameters $\Theta_m^{\mathrm{local}}$, which correspond to the port-selection branch, are kept locally at each base station and are not shared, allowing for flexible adaptation to the specific user distribution and instantaneous interference conditions within each cell.

Then, the personalized unsupervised loss at each BS \(m\) in a given communication round is defined as
\begin{equation}
    \mathcal{L}_m(\Theta_m) = - \sum_{k=1}^K \omega_{m,k} R_{m,k}, 
     \label{eq:22}
\end{equation}
We treat the inter-cell interference \(\hat{I}_{m,k}^{\text{inter}}\) as a constant during each local training round. At the beginning of each global round, it is updated based on the latest global beamforming model and then fixed during local optimization. After local updates, the shared beamforming parameters are aggregated across BSs, and the interference estimation is iteratively updated according to the evolved global beamforming model.

The PFL training executes over $T$ global communication rounds. Within each global round $t$, every participating BS $m$ performs exactly $E$ local iterations (epochs) to iteratively refine both the continuous beamforming and discrete port selection parameters. Let $e \in \{0, 1, \dots, E-1\}$ denote the local iteration index. The gradient descent update rules are formulated as
\begin{subequations}
\begin{align}
    \Theta_m^{\text{local}, (t, e+1)} &= \Theta_m^{\text{local}, (t, e)} - \eta \nabla_{\Theta_m^{\text{local}}} \mathcal{L}_m,
    \label{25a}
    \\
    \Theta_m^{\text{global}, (t, e+1)} &= \Theta_m^{\text{global}, (t, e)} - \eta \nabla_{\Theta_m^{\text{global}}} \mathcal{L}_m,
    \label{25b}
\end{align}
\end{subequations}
Notably, these two parameter sets follow block coordinate descent alternating optimization. First, the global beamforming parameters $\Theta_m^{\mathrm{global}}$ are fixed and the local port selection parameters $\Theta_m^{\mathrm{local}}$ are updated; then $\Theta_m^{\mathrm{local}}$ are fixed and $\Theta_m^{\mathrm{global}}$ are updated. Where $\eta > 0$ represents the learning rate. Guided by these local update rules, the federated workflow in round $t$ proceeds as follows: 

First, BS $m$ reconstructs its complete PA-DNN by combining the downloaded global representation $\Theta^{\text{global}, (t)}$ with its preserved local head $\Theta_m^{\text{local}, (t)}$. The BS then jointly optimizes both parameter sets for $E$ consecutive local epochs via \eqref{25a} and \eqref{25b}. After local training, to strictly preserve the personalized spatial degrees of freedom, the updated discrete port selection branch $\Theta_m^{\text{local}, (t, E)}$ is retained at BS $m$. Conversely, only the updated global representation $\Theta_m^{\text{global}, (t, E)}$ is uploaded to the central server. Finally, the central server performs global aggregation by averaging the uploaded representations from all $M$ BSs to formulate the generalized backbone for the next round:
\begin{equation}
    \Theta^{\text{global}, (t+1)} = \frac{1}{M} \sum_{m=1}^M \Theta_m^{\text{global}, (t, E)}.
\end{equation}

\section{Simulation Results}
\begin{figure}[t]
\centerline{\includegraphics[width=0.45\textwidth]{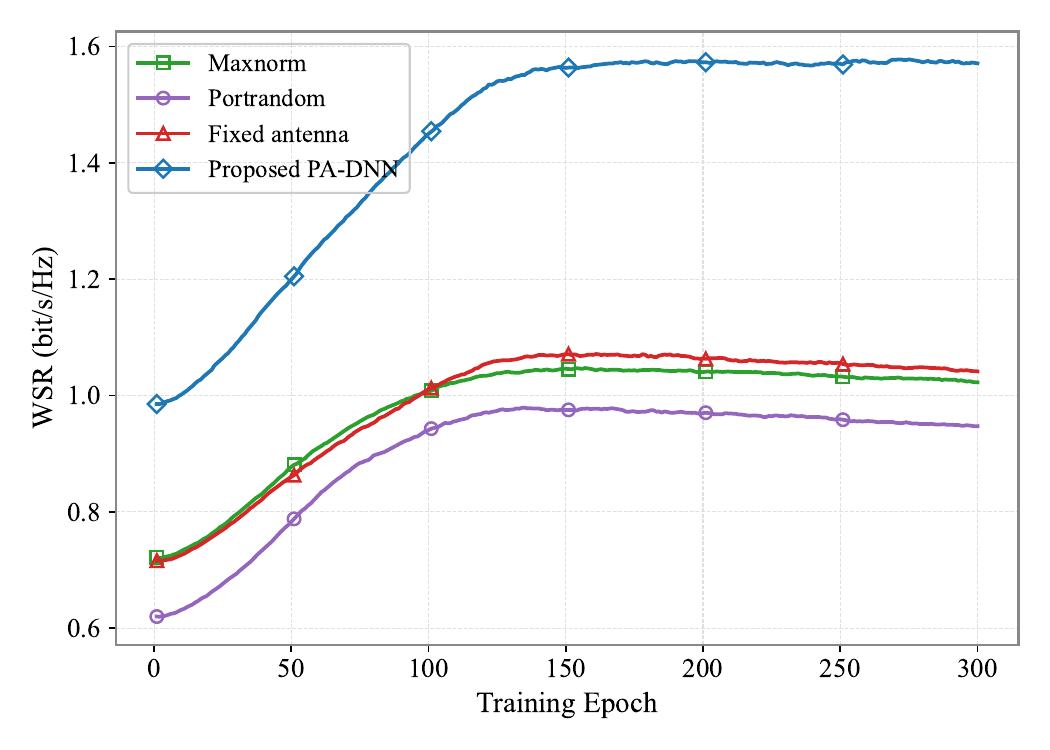}}
\caption{WSR convergence comparison of the proposed PA-DNN against baseline port selection strategies in a multi-cell scenario with \(K = 7\) users (\(K_1 = 2, K_2 = 2, K_3 = 3\)).}
\label{fig:performance_comparison}
\end{figure}

\begin{figure}[t]
\centerline{\includegraphics[width=0.42\textwidth]{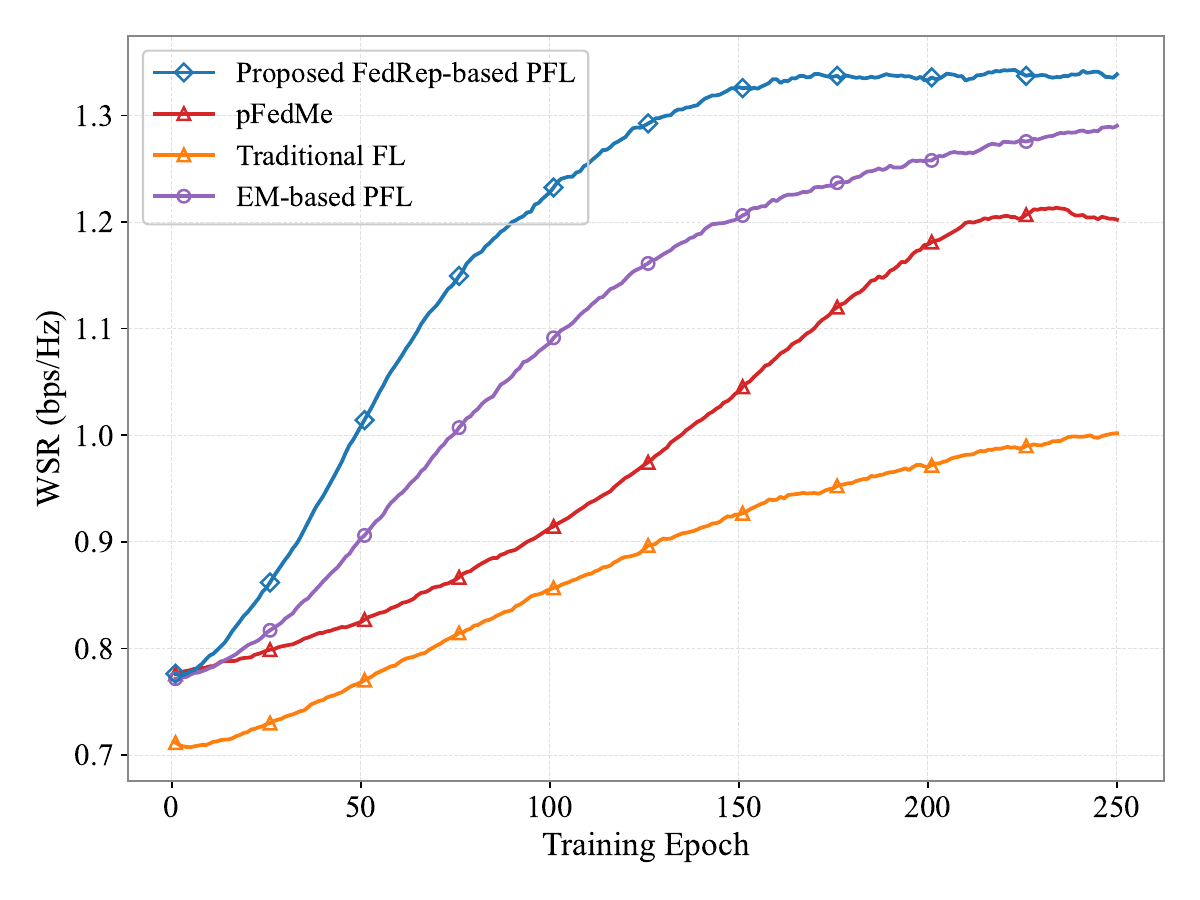}}
\caption{WSR convergence comparison of different federated learning frameworks in a multi-cell scenario with \(K = 9\) users (\(K_1 = 2, K_2 = 3, K_3 = 4\)).}
\label{fig:fedrep_comparison}
\end{figure}

\begin{figure}[t]
\centerline{\includegraphics[width=0.45\textwidth]{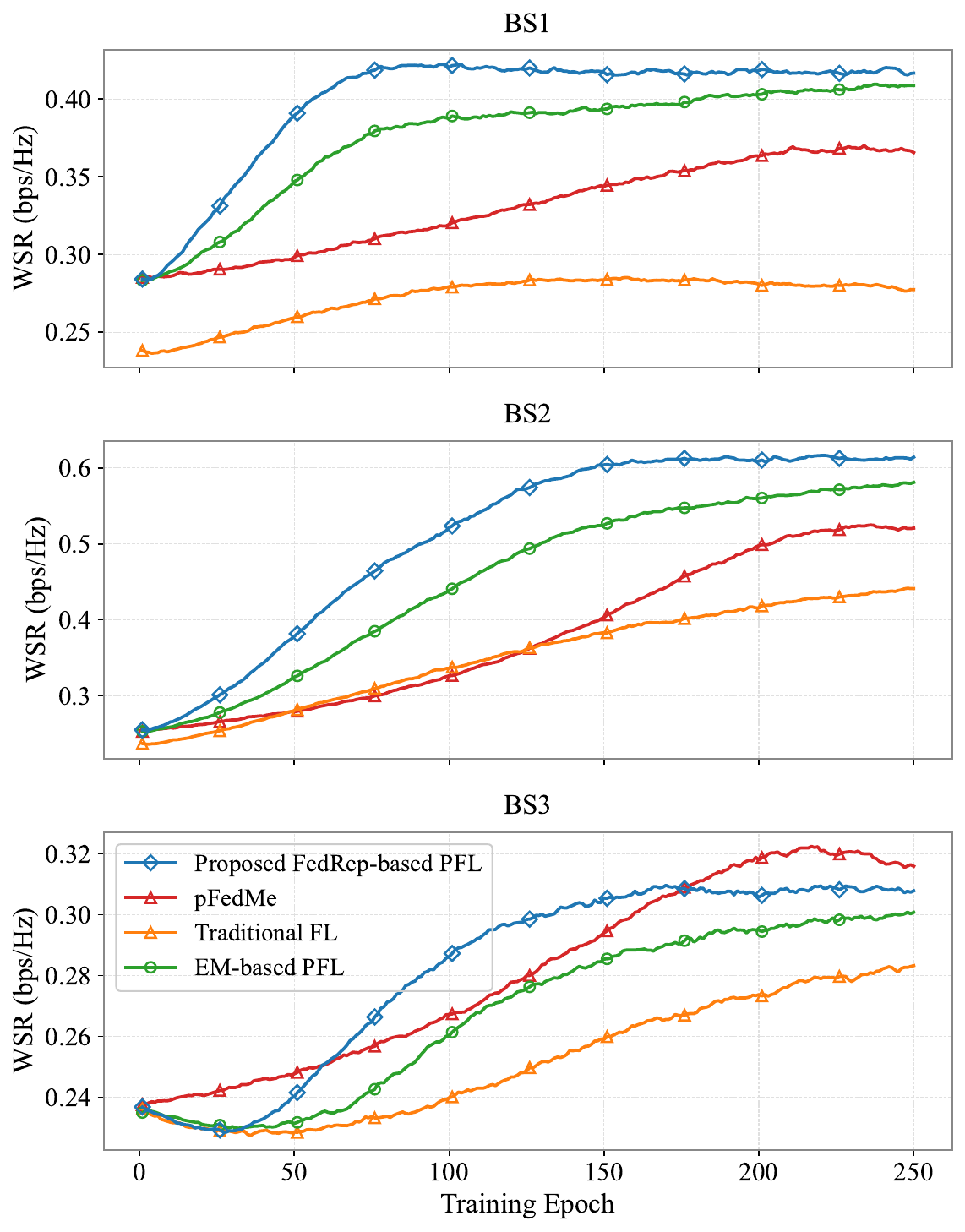}}
\caption{Per-BS WSR performance comparison of different federated learning frameworks in a multi-cell scenario with \(K = 9\) users (\(K_1 = 2, K_2 = 3, K_3 = 4\)).}
\label{fig:per_bs_rate_contrib}
\end{figure}

We consider a downlink transmission at a carrier frequency of $6$ GHz, corresponding to a wavelength $\lambda = 0.05$ m. Each BS is equipped with $N = 8$ transmit antennas, which are assumed to be evenly distributed in a linear antenna array with a spacing of half a wavelength. The maximum transmit power at each BS is set to $30$ dBm. In the following experiments, we adopt two user distribution schemes: $(K_1,K_2,K_3) = (2,2,3)$ in Fig.~\ref{fig:performance_comparison}, and $(2,3,4)$ in Figs.~\ref{fig:fedrep_comparison} and~\ref{fig:per_bs_rate_contrib}. The user weights per BS are set to $[0.6,0.4]$ for two users, $[0.2,0.6,0.2]$ for three users, and $[0.2,0.3,0.2,0.3]$ for four users. Users are uniformly distributed at distances ranging from $15$ m to $60$ m from their serving BS. Each user is equipped with a fluid antenna having $P = 25$ candidate ports located within a $0.02$ m $\times$ $0.02$ m square area. The spatial channels are modeled with $L = 30$ multipath components, free-space path loss (FSPL), and Rayleigh fading. 
The proposed PA-DNN and FedRep frameworks are trained using the Adam optimizer with a batch size of $64$ over $10,000$ generated samples. A learning rate scheduling strategy is employed, where the rate is decayed by a factor of $0.9975$. For FedRep, in each global epoch, both the globally shared beamforming parameters and the locally kept port selection parameters undergo $2$ local updates. Global training terminates either when a predefined number of rounds is reached or when convergence criteria are met.

Fig. \ref{fig:performance_comparison} illustrates WSR convergence of the proposed PA-DNN against various baselines in a multi-cell scenario with $K=7$ users ($K_1=2, K_2=2, K_3=3$). First, the PA-DNN achieves the highest WSR ($1.57$~bps/Hz), significantly outperforming the fixed antenna baseline and demonstrating the fundamental capacity gains enabled by the enhanced spatial degrees of freedom in FASs. Second, its improvement over the Portrandom scheme demonstrates the necessity of intelligent port selection, as random switching fails to exploit spatial correlations. Third, compared with Portrandom, the Maxnorm scheme achieves higher performance during the first 50 epochs, but falls behind after 100 epochs and its growth stagnates. This is because Maxnorm greedily selects locally strongest ports, causing the beamformer to quickly converge to a suboptimal interference-limited solution, while Portrandom encourages broader exploration, enabling better interference suppression over time. Finally, although the PA-DNN requires a slightly longer training period to fully converge (150 epochs), this marginal computational overhead is justified by the steady and robust long-term capacity gains.

Fig. \ref{fig:fedrep_comparison} illustrates WSR performance of the FedRep-based PFL compared against baselines (pFedMe, Traditional FL, EM-based PFL) with user distribution ($K_1=2, K_2=3, K_3=4$). First, the FedRep framework outperforms the benchmark schemes, rapidly converging to the highest WSR ($1.34$~bps/Hz) by decoupling beamforming and discrete port selection into separate parameter modules. Second, Traditional FL forces a common model, ignoring per‑cell data heterogeneity and reaching a plateau early at low capacity ($0.9$~bps/Hz). Third, pFedMe applies $L_2$ regularization and converges after 200 epochs, whereas EM-based PFL employs expectation-maximization to dynamically estimate user-specific weights and achieves the second-highest WSR ($1.29$~bps/Hz), both optimize the variables over the same parameters, causing gradient interference and constraining their performance.

Fig. \ref{fig:per_bs_rate_contrib} illustrates the per-BS WSR performance of the proposed FedRep-based PFL against three baselines, where each BS has a different level of data heterogeneity, with user distribution ($K_1=2, K_2=3, K_3=4$).

\textbf{BS1: Low Heterogeneity and Simple Topology.} 
BS1 serves 2 users with a relatively balanced weight distribution of $[0.6, 0.4]$. In this slightly heterogeneous scenario, all algorithms capture channel conditions, resulting in close performance. Nevertheless, the proposed FedRep achieves the highest WSR ($0.42$ bps/Hz), demonstrating that in simple topologies its decoupled design enables rapid convergence. EM-based PFL follows closely ($0.41$~bps/Hz), as its expectation-maximization blending adapts well to minor weight fluctuations.

\textbf{BS2: High Heterogeneity and Moderate Interference.} 
BS2 serves 3 users with a highly imbalanced weight distribution of $[0.2, 0.6, 0.2]$, so the local optimization objective is heavily dominated by User 2. In contrast to BS1, the proposed FedRep, thanks to its decoupled design, outperforms all baselines under these extreme conditions, achieving the highest WSR ($0.61$~bps/Hz) despite slower convergence due to stronger interference. Its local head remains at each BS, enabling full utilization of spatial degrees of freedom to serve the dominant user without global averaging constraints. Traditional FL and pFedMe intersect twice (around 50 and 120 epochs): Traditional FL initially leads due to fast global convergence, but pFedMe later overtakes it thanks to $L_2$ regularization, achieving a higher final WSR ($0.52$~bps/Hz vs. $0.44$~bps/Hz). EM-based PFL ranks second overall.

\textbf{BS3: Moderate Heterogeneity and Severe Interference.} 
BS3 serves 4 users with a relatively even weight distribution of $[0.2, 0.3, 0.2, 0.3]$. All algorithms experience a brief performance drop because the high Gumbel-Softmax temperature forces nearly random port selection while beamforming is not yet adapted. As temperature decays, port selection becomes deterministic and joint optimization improves. EM-based PFL slightly underperforms FedRep and pFedMe in this dense multi-user scenario with balanced weights, as its EM updates add extra randomness without clear benefit, making convergence slower and less stable. FedRep reaches $0.31$~bps/Hz most quickly by epoch 100, but pFedMe overtakes it around epoch 170 and achieves a slightly higher $0.32$~bps/Hz, because its $L_2$ regularization aids generalization in BS3, allowing continued improvement. Although overtaken, FedRep achieves a good balance: it reaches a competitive rate much earlier (epoch 100) than pFedMe (epoch 170), and the final gap is small ($0.01$~bps/Hz).

\section{Conclusion}
In this paper, we proposed a FedRep-based framework with a PA-DNN to jointly optimize beamforming and port selection in multi-cell FAS. The design decouples global and local parameters, exploits port spatial correlations via positional encoding, and enables decentralized training. Our proposed scheme clearly outperforms traditional baselines, particularly in heterogeneous network environments, confirming the effectiveness of the explicit decoupling and position-aware learning approach.



\bibliographystyle{IEEEtran} 
\bibliography{references1}    

\vspace{12pt}

\end{document}